\documentclass[twocolumn,prl]{revtex4}

\usepackage{graphicx}
\usepackage{amsmath}
\usepackage{hyperref}
\hypersetup{colorlinks,citecolor=black,filecolor=black,linkcolor=black,urlcolor=black}

\makeatletter
\DeclareRobustCommand*{\bfseries}{%
  \not@math@alphabet\bfseries\mathbf
  \fontseries\bfdefault\selectfont
  \boldmath
}
\makeatother

\begin{document}

\title{Analysis of a Continuous-time Model of Structural Balance}

\author{Seth A. Marvel\footnotemark[2]}

\author{Jon M. Kleinberg\footnotemark[3]}
\email{kleinber@cs.cornell.edu}

\author{Robert D. Kleinberg\footnotemark[3]}

\author{Steven H. Strogatz\footnotemark[2]}

\affiliation{\vspace{6 pt} \footnotemark[2]Center for Applied Mathematics, Cornell University, Ithaca, New York 14853 \vspace{1 pt} \\
\footnotemark[3]Department of Computer Science, Cornell University, Ithaca, New York 14853}

\begin{abstract}

It is not uncommon for certain social networks to divide into two opposing camps in response to stress. This happens, for example, in networks of political parties during winner-takes-all elections, in networks of companies competing to establish technical standards, and in networks of nations faced with mounting threats of war. A simple model for these two-sided separations is the dynamical system $dX/dt = X^2$ where $X$ is a matrix of the friendliness or unfriendliness between pairs of nodes in the network.  Previous simulations suggested that only two types of behavior were possible for this system: either all relationships become friendly, or two hostile factions emerge.  Here we prove that for generic initial conditions, these are indeed the only possible outcomes.  Our analysis yields a closed-form expression for faction membership as a function of the initial conditions, and implies that the initial amount of friendliness in large social networks (started from random initial conditions) determines whether they will end up in intractable conflict or global harmony.

\end{abstract}

\maketitle

\section{Introduction}

The mathematical model that we want to study is best understood as an outgrowth of a theory from social psychology known as \textit{structural balance}~\cite{wasserman94}.  So let's begin with a brief explanation of what this theory says.

Consider three individuals:  Anna, Bill and Carl, and suppose that Bill and Carl are friends with Anna, but are unfriendly with each other.  If the sentiment in the relationships is strong enough, Bill may try to strengthen his friendship with Anna by encouraging her to turn against Carl, and Carl might likewise try to convince Anna to terminate her friendship with Bill.  Anna, for her own part, may try to bring Bill and Carl together so they can reconcile and become friends.  In abstract terms, relationship triangles containing exactly two friendships are prone to transition to triangles with either one or three friendships.

Alternately, suppose that Anna, Bill and Carl all view each other as rivals.  In many such situations, there are incentives for the two people in the weakest rivalry to cooperate and form a working friendship or alliance against the third.  In these cases, a single friendship may be prone to appear in a relationship triangle that initially has none.

These two thought experiments suggest a notion of stability, or balance, that can be traced back to the work of Heider~\cite{heider46}.  Heider's theory was expanded into a graph-theoretic framework by Cartwright and Harary~\cite{cartwright56}, who considered graphs on $n$ nodes (representing people, countries or corporations) with edges signed either positive ($+$) to denote friendship or negative ($-$) to denote rivalry.  If a social network feels the proper social stresses (those felt by Anna, Bill and Carl in the examples above), then Cartwright and Harary's theory predicts that in steady state the triangles in the graph should contain an odd number of positive edges---in other words, three positive edges or one positive edge and two negative edges.  We refer to such triangles as \textit{balanced}, and triangles with an even number of positive edges as \textit{unbalanced}.  Finally, we call a graph {\em complete} if it contains edges between all pairs of nodes, and we say that a complete graph with signs on its edges is \textit{balanced} if all its triangles are balanced.  (All graphs in our discussion will be complete.)

As it turns out, these local notions of balance theory are closely related to the global structure of two opposing factions.  In particular, suppose that the nodes of a complete graph are partitioned into two factions such that all edges inside each faction are positive and all edges between nodes in opposite factions are negative.  (One of these factions may be empty, in which case the other faction includes all the nodes in the graph, and consequently all edges of the network are positive.)  Note that this network must be balanced, since each triangle either has all three members in the same faction (yielding three positive edges) or has two members in one faction and the third member in the other faction (yielding one positive edge and two negative ones).  In fact, a stronger and less obvious statement is true:  any balanced graph can be partitioned into two factions in this way, with one faction possibly empty~\cite{cartwright56}.  As a result, when we speak of balanced graphs, we can equivalently speak of networks with this type of two-faction structure.

\section{Model}

Structural balance is a static theory---it posits what a ``stable'' signing of a social network should look like.  However its underlying motivation is dynamic, based on how unbalanced triangles ought to resolve to balanced ones.  This situation has led naturally to a search for a full dynamic theory of structural balance.  Yet finding systems that reliably guide networks to balance has proved a challenge in itself.

A first exploration of this issue was conducted by Antal et al.~\cite{antal05} who considered a family of discrete-time models.  In one of the main models of this family, an edge of the graph is examined in each time step, and its sign is flipped if this produces more balanced triangles than unbalanced ones.  While a balanced graph is a stable point for these discrete dynamics, it turns out that many unbalanced graphs called \textit{jammed states} are as well~\cite{antal05,marvel09}.

Thus, the natural problem became to identify and rigorously analyze a simple system that could progress to balanced graphs from generic initial configurations.  A novel approach to this problem was taken by Ku{\l}akowski, Gawro{\'n}ski, and Gronek~\cite{kulakowski05}, who proposed a continuous-time model for structural balance.  They represented the state of a completely connected social network using a real symmetric $n \times n$ matrix $X$ whose entry $x_{ij}$ represents the strength of the friendliness or unfriendliness between nodes $i$ and $j$ (a positive value denotes a friendly relationship and a negative value an unfriendly one).  Note that for a given $X$, there is a signed complete graph with edge signs equal to the signs of the corresponding elements $x_{ij}$ in $X$.  We will call $X$ balanced if this associated signed complete graph is balanced.

Ku{\l}akowski et al. considered variations on the following basic differential equation, which they proposed as a dynamical system governing the evolution of the relationships over time:

	\begin{equation} \label{1}
	\frac{dX}{dt} = X^2.
	\end{equation}

\noindent Remarkably, simulations showed that for essentially any initial $X(0)$, the system reached a balanced pattern of edge signs in finite time.  

Writing Eq.~\ref{1} directly in terms of the entries $x_{ij}$ gives a sense for why this differential equation should promote balance:

	\begin{equation} \label{2}
	\frac{dx_{ij}}{dt} = \sum_k x_{ik} x_{kj}.
	\end{equation}

\noindent Notice that $x_{ij}$ is being pushed in a positive or negative direction based on the relationships that $i$ and $j$ have with $k$:  if $x_{ik}$ and $x_{kj}$ have the same sign, their product guides the value of $x_{ij}$ in the positive direction, while if $x_{ik}$ and $x_{kj}$ have opposite signs, their product guides the value of $x_{ij}$ in the negative direction.  In each case, this is the direction required to balance the triangle $\{i,j,k\}$.  Note also that Eq.~\ref{2} applies for the case that $i = j$.  While this case is harder to interpret, the monotonic increase of $x_{ii}$ implied by Eq.~\ref{2} might be viewed in psychological terms as an increase of self-approval or self-confidence as $i$ becomes more resolute in its opinions about others in the network.

For a network with just three nodes, it can be easily proved that a variant of these dynamics generically balances the single triangle in this network; such a three-node analysis has been given by Ku{\l}akowski et al.~\cite{kulakowski05}, and we describe a short proof in the Supporting Information.  What is much less clear, however, is how the system should behave with a larger number of nodes, when the effects governing any one edge $\{i,j\}$ are summed over all nodes $k$ to produce a single aggregate effect on $x_{ij}$.

It has therefore been an open problem to prove that Eq.~\ref{2} or any of the related systems studied by Ku{\l}akowski et al. will bring a generic initial matrix $X(0)$ to a balanced state.  It has also been an open problem to characterize the structure of the balanced state that arises as a function of the starting state $X(0)$.

\section{Results}

In this paper, we resolve these two open problems.  We first show that for a random initial matrix (drawn from any absolutely continuous distribution), the system reaches a balanced matrix in finite time with a probability converging to $1$ in the number of nodes $n$.  In addition, we provide a closed-form expression for this balanced matrix in terms of the initial one; essentially, we discover that the system of differential equations serves to ``collapse'' the starting matrix to a nearby rank-one matrix.  We also characterize additional aspects of the process, giving for example a description of an ``exceptional'' set of matrices of probability measure converging to $0$ in $n$ for which the dynamics are not necessarily guaranteed to produce a balanced state.

We then analyze the solutions of the system for classes of random matrices in the large-$n$ limit---in particular, we consider the case in which each unique matrix entry is drawn independently from a distribution with bounded support that is symmetric about a number $\mu$ (the mean value of the initial friendliness among the nodes).  In this case, we find a transition in the solution as $\mu$ varies:  when $\mu > 0$, the system evolves to an all-positive sign pattern, whereas when $\mu \leq 0$, the system evolves to a state in which the network is divided evenly into two all-positive cliques connected entirely by negative edges.  We end by discussing some implications of the model and the associated transition between harmony and conflict, including an evaluation of the model on empirical data and some potential connections to research on reconciliation in social psychology.

\subsection{Behavior of Model:  Evolution to a Balanced State}

Suppose we randomly select the $x_{ij}(0)$'s from a continuous distribution on the real line.  Then the $x_{ij}(t)$'s found by numerical integration generally sort themselves in finite time into the sign pattern of two feuding factions.  To reformulate this observation as a precise statement and explain why the behavior holds so pervasively, we now solve Eq.~\ref{1} explicitly.

\vspace{4pt}

\textbf{Solution to model.}  The initial matrix $X(0)$ is real and symmetric by assumption, so we can write it as $Q D(0) Q^T$ where $D(0)$ is the diagonal matrix with the eigenvalues of $X(0)$, denoted $\lambda_1 \geq \lambda_2 \geq \cdots \geq \lambda_n$, as diagonal entries ordered from largest to smallest, and $Q$ is the orthogonal matrix with the corresponding eigenvectors of $X(0)$, denoted $\omega_1, \omega_2, \dotsc, \omega_n$, as columns.  The superscript $T$ signifies transposition.

The differential equation Eq.~\ref{1} is a special case of a general family of equations known as \textit{matrix Riccati equations} \cite{aboukandil03}.  The analysis of the full family is complicated and not fully resolved, but we now show that the special case of concern to us, Eq.~\ref{1}, has an explicit solution with a form that exposes its connections to structural balance.  We proceed as follows.  First, we observe that by separation of variables, the solution of the single-variable differential equation $\dot{x} = x^2$ (overdot representing differentiation by time) with initial condition $x(0) = \lambda_k$ is

	\begin{equation} \label{3}
	\ell_k(t) = \frac{\lambda_k}{1 - \lambda_k t}.
	\end{equation}
	
\noindent Therefore the diagonal matrix $D(t) = {\rm diag}(\ell_1(t),$ $\ell_2(t), \dotsc,$ $\ell_n(t))$ is the solution of Eq.~\ref{1} for the initial condition $X(0) = {\rm diag}(\lambda_1, \lambda_2, \dotsc, \lambda_n)$.

Moreover $Y(t) = Q D(t) Q^T$ is also a solution of Eq.~\ref{1} since $\dot{Y} = Q \dot{D} Q^T = Q (D^2) Q^T = (Q D Q^T)^2 = Y^2$.  But $Y(t)$ has the same initial condition as $X(t)$ in our original problem:  $Y(0) = Q D(0) Q^T = X(0)$.  So by uniqueness, $Y(t) = Q D(t) Q^T$ must be the solution we seek.

Our solution $X(t)$ can also be written in a different way to mimic the solution of the one-dimensional equation $\dot{x} = x^2$.  Since $x_{ij}(t) = \sum_{k = 1}^n q_{ik} \ell_k(t) q_{jk}$, where $q_{ij}$ is the $(i,j)$th entry of $Q$, we can expand the denominators of the $\ell_k(t)$ functions in powers of $t$ to rewrite $X(t)$ as $X(0) + X(0)^2 t + X(0)^3 t^2 + \dotsb$, or more concisely,

	\begin{equation} \label{4}
        X(t) = X(0) [I - t X(0)]^{-1}.
	\end{equation}
	
\noindent (Note that the matrices $X(0)$ and $[I - X(0) t]^{-1}$ commute.)  This equation is valid when $t$ is less than the radius of convergence of every $\lambda_k$, that is when $t < 1/\lambda_1$ (assuming $\lambda_1 > 0$).

Finally we note that the above method of solving Eq.~\ref{1} contains a reduction of the number of dynamical variables of the system from $\binom{n+1}{2}$ to $n$.  The $\binom{n}{2}$ constants of motion generated by this reduction are just the off-diagonal elements of $Q^T X(t) Q = D(t)$, or $\sum_{k = 1}^n \sum_{\ell = 1}^n q_{ki} x_{k\ell}(t) q_{{\ell}j} = 0$ for all $1 \leq i < j \leq n$. Furthermore, the procedure for reducing $X(t)$ can be easily generalized to any system of the form $\dot{X} = f(X)$ where $f$ is a polynomial of $X$.

\vspace{4pt}

\textbf{Behavior of solution.}  Let's now examine the behavior of our solution $X(t)$ to see why in the typical case it splits into two factions in finite time.  It turns out that this is the guaranteed outcome if the following three conditions hold (and as we will see below, they hold with probability converging to $1$ as $n$ goes to infinity):
\vspace{4pt}

1. $\lambda_1 > 0$,

2. $\lambda_1 \neq \lambda_2$ (and hence $\lambda_1 > \lambda_2$), and

3. all components of $\omega_1$ are nonzero. \vspace{4pt}

\noindent To see why these conditions imply a split into two factions, observe from Eq.~\ref{3} that each $\ell_k(t)$ diverges to infinity at $t = 1/\lambda_k$.  Since $x_{ij}(t) = \sum_{k = 1}^n q_{ik} \ell_k(t) q_{jk}$, all $x_{ij}$'s diverge to infinity when the $\ell_k$ with the smallest positive $1/\lambda_k$ does.  Under the first and second conditions, this $\ell_k$ is $\ell_1$, so the blow-up time $t^*$ of Eq.~\ref{1} must be $1/\lambda_1$.  To show that the nodes are partitioned into two factions as $X(t)$ approaches $t^*$, let $\overline{X}(t) = X(t) / ||X(t)||$ on the half-open interval $[0, t^*)$, where $||X(t)||$ denotes the Frobenius norm of $X$. The matrix $\overline{X}(t)$ has the sign pattern of $X(t)$, and as $t$ approaches $t^*$ it converges to the rank-one matrix

	\begin{equation} \label{5}
	X^* = Q \; {\rm diag}(1, 0, 0, \dotsc, 0) \; Q^T = \omega_1 \omega_1^T
	\end{equation}

\noindent Now let $\omega_{1k}$ denote the value of the $k{\rm th}$ coordinate of $\omega_1$, and let $S = \{k : \omega_{1k} > 0\}$ and $T = \{k : \omega_{1k} < 0\}.$  Then $S$ and $T$ partition the node indices $1, 2, \dotsc, n$ by our condition that $\omega_1$ has no zero components.  From Eq.~\ref{5}, this partition must correspond to two cliques of friends joined by a complete bipartite graph of unfriendly ties.

\vspace{4pt}

\textbf{The three conditions.}  We now return to the three conditions above.  We first show that the second and third hold with probability $1$.  We then show that the first condition holds with probability converging to $1$ as $n$ goes to infinity.  Lastly, we analyze the behavior of the system in the unlikely event that the first condition does not hold.  The fact that the conjunction of all three conditions holds with probability converging to $1$ as $n$ grows large justifies our earlier claim that the behavior described above holds for almost all choices of initial conditions.

First we show why the second and third conditions hold with probability 1 so long as the (joint) distribution from which $X(0)$ is drawn is absolutely continuous with respect to Lebesgue measure---in other words, assigns probability zero to any set of matrices whose Lebesgue measure is zero.  Our arguments below make use of the following two basic facts:
\vspace{4pt}

i. the set of zeros of a nontrivial multivariate polynomial has Lebesgue measure zero, and

ii. the existence of a common root of two univariate polynomials $P$ and $Q$ is equivalent to the vanishing of a multivariate polynomial in the coefficients of $P$ and $Q$ (specifically, it is equivalent to the vanishing of the determinant of the Sylvester matrix of $P$ and $Q$, also called the \textit{resultant} of $P$ and $Q$). \vspace{4pt}

To show that $\lambda_1 \neq \lambda_2$ with probability 1, let $P$ denote the characteristic polynomial of $X(0)$, and let $Q$ denote the derivative of $P$.  Then $X(0)$ has a repeated eigenvalue if and only if $P$ has a repeated root, which it does if and only if $P$ and $Q$ have a common root.  This condition is equivalent to the vanishing of the resultant of $P$ and $Q$, which is a multivariate polynomial in the entries of $X(0)$.  The polynomial cannot be zero everywhere, because there is at least one symmetric matrix that does not have a repeated eigenvalue.  So the set of matrices having a repeated eigenvalue has Lebesgue measure zero.

	\begin{figure*}[t]
	\centering
	\includegraphics[scale = 1]{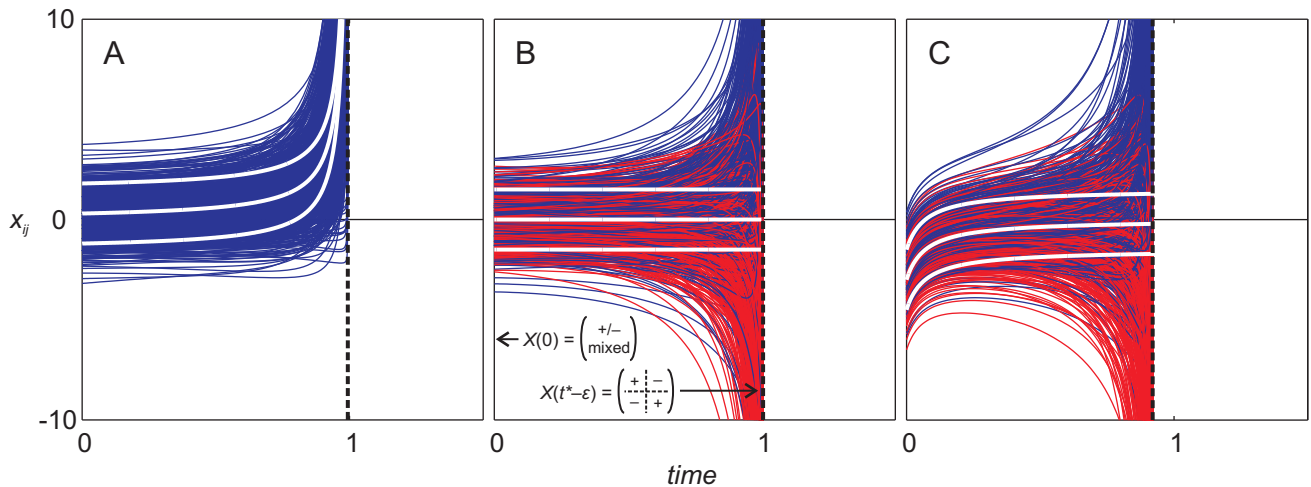}
	\caption{Representative large-$n$ plots of the model for (A) $\mu > 0$ ($\mu = 3/10$ in the plot shown), (B) $\mu = 0$, and (C) $\mu < 0$ ($\mu = -3$ in the plot shown).  For all three plots, $\sigma = 1$ and $n = 90$.  To reduce image complexity, only one randomly sampled fifth of the trajectories is included.  In the second plot, $t^*$ denotes the time at which the system diverges, and $\epsilon$ denotes a sufficiently small displacement.  The white curves superimposed on the three plots are the large-$n$ trajectories $x_{ij}(t)= x_{ij}(0) - \mu + \mu/(1 - \mu nct)$ for $x_{ij}(0) = \mu, \mu \pm 3\sigma/2$, where $c$ represents a rescaling of time.  Since we want to fix the blow-up time $t^*$ near 1 and since $ct^* = 1/\lambda_1$ as found in the text, we choose $c = 1/(\mu n + \nu - \mu + \sigma^2/\mu)$ for (A) and $c = 1/(2\sigma\sqrt{n})$ for (B) and (C) using estimates of $\lambda_1$ taken from Ref.~\cite{furedi81}.  The black dotted lines mark the blow-up times $t^* = 1/(c\lambda_1)$. \label{largenregimes}}
	\end{figure*}

Similarly, to show that all components of $\omega_1$ are nonzero, let $P$ denote the characteristic polynomial of $X(0)$ and $P_i$ the characteristic polynomial of the $(n-1) \times (n-1)$ submatrix $X_i(0)$ obtained by deleting the $i$th row and $i$th column of $X(0)$.  It is easy to check that if any eigenvector of $X(0)$ has a zero in its $i$th component, then the vector obtained by deleting that component is an eigenvector of $X_i(0)$ with the same eigenvalue.  Consequently, $P$ and $P_i$ must have a common root, implying that the resultant of $P$ and $P_i$ vanishes.  This resultant is once again a multivariate polynomial in the entries of $X(0)$, and once again it must be nonzero somewhere because there is at least one symmetric matrix whose eigenvectors all have nonzero entries.  Hence, the set of matrices having an eigenvector with zero in its $i$th component has Lebesgue measure zero.

Finally, to determine the likelihood of the first condition, we first must say a bit more about the way that $X(0)$ is selected.  Suppose that the off-diagonal $x_{ij}(0)$'s are drawn randomly from a common distribution $F$ and the on-diagonal $x_{ii}(0)$'s are drawn randomly from a common distribution $G$.  All selections are independent for $i \leq j$.  (For $i > j$, we let $x_{ij}(0) = x_{ji}(0)$, so that $X(0)$ is symmetric.) For this construction of $X(0)$, Arnold~\cite{arnold71} has shown that with the remarkably weak additional assumption that $F$ has a finite second moment, Wigner's semicircle law holds in probability as $n$ grows to infinity.  This in turn implies that $\lambda_1 > 0$ in probability in the same limit.

Moreover, suppose we are in the low-probability case that $\lambda_1 \leq 0$.  In this case, the analysis above shows that all the functions $\ell_i(t)$ converge to $0$ as $t \rightarrow \infty$.  Thus, $\lim_{t \rightarrow \infty} D(t) = 0$, and since $X(t) = Q D(t) Q^T$, we also have $\lim_{t \rightarrow \infty} X(t) = 0.$

Although the entries of $X(t)$ converge to zero when $\lambda_1 \leq 0$, one might still want to know if the sign pattern of $X(t)$ is eventually constant (i.e., remains unchanged for all $t$ above some threshold value) and, if so, what determines this sign pattern.  It is possible to answer this question, again assuming the second and third conditions.  By expanding the function $\ell_i(t) = \lambda_i/(1 - \lambda_i t)$ in powers of $u = 1/t$, we obtain the asymptotic series

	\begin{equation} \label{7}
	\ell_i(t)  =  -u - u^2 \lambda_i^{-1} - O(u^3),
	\end{equation}
	
\noindent which implies

	\begin{equation} \label{8}
	X(t) = Q D(t) Q^T = -uI - u^2 X(0)^{-1} - O(u^3).
	\end{equation}

\noindent In the limit of small $u$, the leading order term of the diagonal entries of $X(t)$ is the linear term, which has negative sign.  For the off-diagonal entries of $X(t)$, the leading-order term as $u$ tends to zero is the quadratic term, whose sign matches the sign of the corresponding off-diagonal entry of the matrix $-X(0)^{-1}$.

\subsection{Behavior of Model:  From Factions to Unification}

The analysis in the previous section tells us how to find both the blow-up time $t^*$ and final sign configuration of a network if we know its initial state $X(0)$.  However we might also want to know whether we can characterize the behavior of $X(t)$ in the large-$n$ limit in terms of statistical parameters of $X(0)$.  This could, for example, help us forecast the behavior of large populations when collecting complete relationship-level data is not feasible.  

In this section, we show that there is a transition from final states consisting of two factions to final states consisting of all positive relations as the ``mean friendliness'' of $X(0)$ (the mean of the distributions used to generate the off-diagonal entries of $X(0)$) is increased from negative to positive values.  This is consistent with the numerical simulations shown in Fig.~\ref{largenregimes}.

Before discussing the details though, we should describe how $X(0)$ is selected in this section.  We start by adopting the procedure of F{\"u}redi and Koml{\'o}s~\cite{furedi81}:  the elements $x_{ij}(0)$ are drawn independently from distributions $F_{ij}$ with zero mass outside of $[-K, K]$.  The off-diagonal $F_{ij}$'s have a common expectation $\mu$ and finite variance $\sigma^2$, while the on-diagonal $F_{ii}$'s have a common expectation $\nu$ and variance $\tau^2$.  In addition, we require that each off-diagonal distribution $F_{ij}$ be symmetric about $\mu$.  Now let's consider the three cases of positive, zero and negative $\mu$.

\vspace{4pt}

\textbf{Case 1: {\normalsize $\mu > 0$}.}  The results of F{\"u}redi and Koml{\'o}s~\cite{furedi81} show that when $\mu > 0$, the deviation of $\omega_1$ from $(1, 1, \dotsc, 1)/\sqrt{n}$ vanishes in probability in the large-$n$ limit.  Hence the final state of the system consists of one large clique of friends containing all but at most a vanishing fraction of the nodes.  Moreover, by assuming a bound on $\sigma$ we can strengthen this statement further: if $\sigma < \mu/2$, then the findings of F{\"u}redi and Koml{\'o}s imply that the final state consists of a single clique of friends, with no negative edges.  These observations are consistent with the representative numerical trial shown in Fig.~\ref{largenregimes}A.  Moreover, F{\"u}redi and Koml{\'o}s show that the asymptotic behavior of $\lambda_1$ grows like $\mu n + O(1)$, and hence the blow-up time scales like $1/(\mu n)$.

We can gain insight into the behavior of the system for small $t$ using an informal Taylor series calculation:  if we rescale time in Eq.~\ref{1} by inserting a $1/n$ before the summation, compute the Taylor expansion of $x_{ij}(t)$ term-by-term and then take the expectation of each term, we obtain the geometric series $x(t) = \mu + \mu^2 t + \mu^3 t^2 + \dotsb$, or

	\begin{equation} \label{9}
	x(t) = \frac{\mu}{1 - \mu t}.
	\end{equation}

\noindent With significantly more work, it can be proved that every trajectory $x_{ij}(t)$ has this time dependence on $[0,1/K)$ in the large-$n$ limit with probability 1 (see the Supporting Information), so we may write

	\begin{equation} \label{10}
	\lim_{n \rightarrow \infty} x_{ij}(t)
	= x_{ij}(0) - \mu + \frac{\mu}{1 - \mu t} \quad \text{with prob. 1}
	\end{equation}

\noindent for all $t$ in $[0,1/K)$.  Observe that this limit has a blow-up time $t^*$ of $1/\mu$.  Since our rescaling of time represents a zooming in or magnification of time by a factor of $n$, this $t^*$ corresponds to a blow-up time asymptotic to $1/(\mu n)$ for the unrescaled system, consistent with the results of F{\"u}redi and Koml{\'o}s.

\vspace{4pt}

\textbf{Case 2: {\normalsize $\mu = 0$}.}  In the event that the network starts from a mean friendliness of zero, numerical experiments indicate that the system ends up with two factions of equal size in the large-$n$ limit (Fig.~\ref{largenregimes}B).  We now prove this to be the case.  For the remainder of this discussion, we will abbreviate $X(0)$ as $A$ and $x_{ij}(0)$ as $a_{ij}$.

Since the off-diagonal entries of $A$ have symmetric distributions by assumption, we have for any off-diagonal $a_{ij}$ and any interval $S_{ij}$ on the real line that $P(a_{ij} \in S_{ij}) = P(-a_{ij} \in S_{ij})$.  Now let $D$ be a diagonal matrix with some sequence of $+1$ and $-1$ along its diagonal (where the $i$th diagonal entry is denoted by $d_i$).  Then the random matrices $A$ and $B = DAD$ are identically distributed, as we will now show.

To say that $A$ and $B$ are identically distributed means that for every Borel set of matrices $S$, $P(A \in S) = P(B \in S)$.  To prove this, it suffices to consider the case in which $S$ is a product of intervals $S_{ij}$, since these product sets generate the Borel sigma-algebra.  The entries of $A$ are independent, so $P(A \in S) = \Pi_{i \leq j} P(a_{ij} \in S_{ij})$.  Similarly, $P(B \in S) = \Pi_{i \leq j} P(d_i a_{ij} d_j \in S_{ij})$.  By the symmetry of the off-diagonal distributions, $\Pi_{i \leq j} P(a_{ij} \in S_{ij}) = \Pi_{i \leq j} P(d_i a_{ij} d_j \in S_{ij})$, which gives us $P(A \in S) = P(B \in S)$ as desired.  (Note that when $i = j$, the factor $d_id_j$ is 1 so the on-diagonal distributions need not be symmetric.)
	
Now consider the set $S$ of matrices with an $\omega_1$ consisting of all positive components.  The above demonstration implies that the probability of choosing an $A$ in this set is the same as choosing an $A$ such that $B$ is in this set.  Regarding the later event, $A(D\omega_i) = \lambda_i(D\omega_i)$ implies $B\omega_i = \lambda_i\omega_i$, so the $\lambda_1$ eigenvector of the $A$ used to compute $B$ is $D\omega_1$.  This demonstrates that all sign patterns for the components of $\omega_1$ are equally likely.  In other words, the distribution of the number of positive components in $\omega_1$ is the binomial distribution $B(n,1/2)$ and the fraction of positive components in $\omega_1$ converges (in several senses) to $1/2$ as $n$ grows large.

Additionally, we can consider how $\lambda_1$ varies with $n$ in the case that $\mu = 0$ to determine when the blow-up will occur.  F{\"u}redi and Koml{\'o}s~\cite{furedi81} found for this case that $\lambda_1 \in 2 \sigma \sqrt{n} + O(n^{1/3}\log n)$ with probability tending to 1, so with probability tending to 1 the blow-up time shrinks to zero like $1/\sqrt{n}$, an order of $\sqrt{n}$ slower than in the $\mu > 0$ case.

\vspace{4pt}

\textbf{Case 3: {\normalsize $\mu < 0$}.}  For this final case, F{\"u}redi and Koml{\'o}s~\cite{furedi81} found that $\lambda_1 < 2 \sigma \sqrt{n} + O(n^{1/3}\log n)$ with probability tending to 1.  The semicircle law gives a lower bound:  $\lambda_1 > 2 \sigma \sqrt{n} + o(\sqrt{n})$ in probability.  So the blow-up time goes to zero like $1/\sqrt{n}$ in the unrescaled system.

Note also that if we define a new matrix $C = -A$ where $A$ is now the initial matrix $X(0)$ of Case 3, then $C$ satisfies the condition of Case 1, $\mu > 0$.  Thus the distance between the top eigenvector of $C$ and $(1, 1, \dotsc, 1)/\sqrt{n}$ declines to zero in probability just as in Case 1.  Furthermore, every other eigenvector of $C$ is orthogonal to the largest one.  Hence if $\sigma < |\mu|/2$, then with probability tending to $1$, every other eigenvector acquires a mixture of positive and negative components in the large-$n$ limit, including the bottom eigenvector of $C$, which is the top eigenvector of $A$.  This establishes that in the case that $\mu < 0$ and $\sigma < |\mu|/2$, the system ends up in a state with two factions with probability converging to $1$ for all finite $n$.

Numerical simulations of the case that $\mu < 0$ suggest the conjecture that the two factions are approximately equal in size for large $n$.  Furthermore, the derivation of Eq.~\ref{10} is in fact valid for all $\mu$, so each trajectory rapidly decays from $x_{ij}(0)$ toward $x_{ij}(0) - \mu$ on $[0,1/K)$ (Fig.~\ref{largenregimes}C).  This transient decay appears to extend beyond $t = 1/K$ in numerical simulations.  So, for example, if time is rescaled by $1/\sqrt{n}$ instead of $1/n$, we would hypothesize that (i) each trajectory makes a complete jump from $x_{ij}(0)$ to $x_{ij}(0) - \mu$ in the large-$n$ limit, and that (ii) from this point onward, the system behaves like an initial configuration of the $\mu = 0$ case and so separates into two equal factions \textit{en route} to its blow-up at $1/(2\sigma)$.

	\begin{figure*}[t]
	\centering
	\includegraphics[scale = 1]{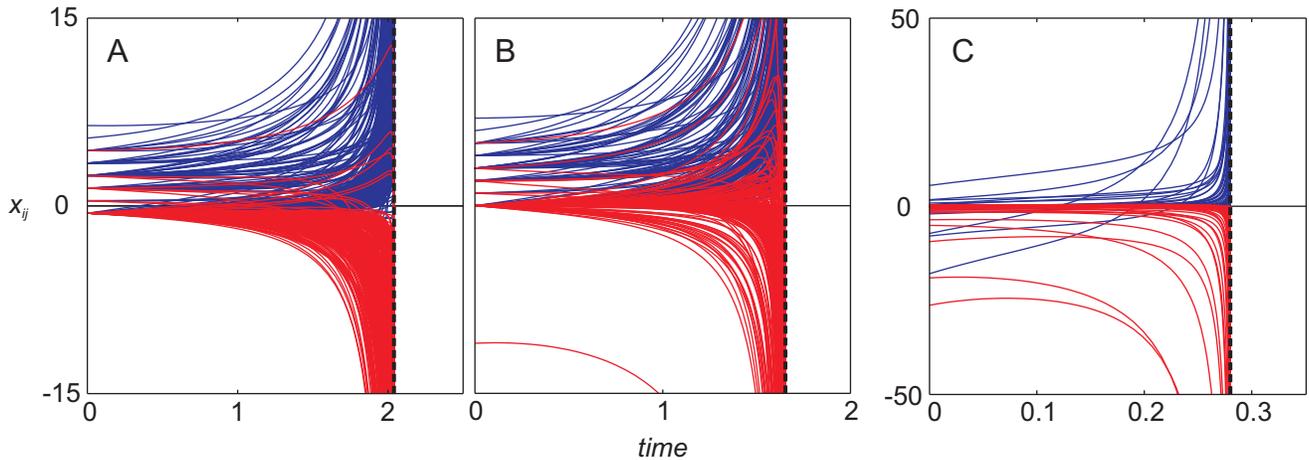}
	\caption{Tests of the model of Ku{\l}akowski et al. (Eq.~\ref{1}) against two existing data sets.  (A) The evolution of the model starting from Zachary's capacity matrix with the capacity of each relationship reduced by 0.58.  This is the minimal downward displacement necessary (to two significant figures) for the resulting separation to be correct for all but 1 of the 34 club members.  For reasons described by Zachary~\cite{zachary77}, this is basically the best separation we can expect.  (B) The evolution of the model from Zachary's capacity matrix with the capacity of zero between the two club leaders replaced by $-11$; the resulting factions are identical to those in (A).  Substituted values less than $-11$ yield the same two factions, while greater values produce less accurate divisions.  (C) The evolution of the model starting from Axelrod and Bennett's 1939 ${\rm propensity}(i,j) \cdot {\rm size}(i) \cdot {\rm size}(j)$ matrix for the 17 countries involved in World War II (by Axelrod and Bennett's definition).  The model finds the correct split into Allied and Axis powers with the exceptions of Denmark and Portugal.  Axelrod and Bennett's own landscape theory of aggregation does slightly better---its only misclassification is Portugal. \label{datacomparison}}
	\end{figure*}
	
\section{Discussion}

In this final section, we review our results and their significance relative to previous work in structural balance theory.  We then compare the predictions of the model with data, discuss potential criticisms of the model, and finish with some intriguing connections between the behavior of the model and recent social-psychological work on neutralizing two-sided conflicts.

Our first result is a demonstration that the model forms two factions in finite time across a broad set of initial conditions.  As noted at the outset, similar demonstrations have not been possible for dynamic models of structural balance in earlier literature because these models contained so-called \textit{jammed states} that could trap a social network before it reached a two-faction configuration~\cite{antal05,marvel09}.  The model of Ku{\l}akowski et al. by contrast has no such jammed states for generic initial conditions and hence provides a robust means for a social network to balance itself.

The second result of the paper is the discovery and characterization of a transition from global polarization to global harmony as the initial mean friendliness of the network crosses from nonpositive to positive values.  Similar transitions have been observed in other models of structural balance but so far none has been characterized at a quantitative level.  For example, Antal et al.~\cite{antal05} found a nonlinear transition from two cliques of equal size to a single unified clique as the fraction of positively signed edges at $t = 0$ was increased from 0 to 1 (see Fig. 5 of Ref.~\cite{antal05}).  The authors provided a qualitative argument for this transition, but left open the problem of its quantitative detail.  Our results both confirm the generality of their observations and provide a quantitative account of a transition analogous to theirs.

To complement the theoretical nature of our work and get a better sense of how the model behaves in practice, we can numerically integrate it for several cases of empirical social network data where the real-life outcomes of the time-evolution are known.  Our first example is based on a study by Zachary~\cite{zachary77} who witnessed the break-up of a karate club into two smaller clubs.  Prior to the separation, Zachary collected counts of the number of social contexts in which each pair of individuals interacted outside of the karate club, with the idea being that the more social contexts they shared, the greater the likelihood for information exchange.  These counts, or \textit{capacities} as Zachary called them, can be converted to estimates of friendliness and rivalry in many different ways.  For a large class of such conversions, Eq.~\ref{1} predicts the same division that Zachary's method found, which misclassified only 1 of the 34 club members (Fig.~\ref{datacomparison}A,B).  

A second example can be constructed from the data of a study by Axelrod and Bennett~\cite{axelrod93} regarding the aggregation of Allied and Axis powers during World War II.  If we simply take the entries of their ${\rm propensity}(i,j) \cdot {\rm size}(i) \cdot {\rm size}(j)$ matrix to be proportional to the friendliness felt between the various pairs of countries in the war, then running the model gives the correct Allied-Axis split for all countries except Denmark and Portugal (Fig.~\ref{datacomparison}C).

Despite these modest successes, the model could still be criticized as ``a simplification and an idealization, and consequently a falsification'' \cite{turing52}.  Clearly, human behavior is more complicated than what is captured by Eq.~\ref{1}.  However, deliberate simplicity is a common feature of many foundational mathematical models of basic social phenomena, which are often designed to isolate and study the effect of a single social force.  Such models can be particularly appropriate in extreme settings where this single force plays a dominant role, making human choices more constrained and thus perhaps more predictable.  In the present case, the Ku{\l}akowski et al. model is designed to ignore all other social behaviors besides the urge to make one's friendships and rivalries consistent.  In this respect it is a bit like problems in classical physics involving frictionless surfaces and massless springs; it is a mathematical cartoon of a single aspect of our social experience.  It may give mechanistic insight but is not designed for quantitative prediction.

A more specific objection might be raised regarding the divergence to infinity in finite time.  However, since the purpose of the model is to study the pattern of signs that emerges, our main conclusion from the model is that the sign pattern eventually stabilizes at a point before the divergence.  This stabilization of the sign pattern is our primary focus, and one could interpret the subsequent singularity as simply the straightforward and unimpeded ``ramping up'' of values caused by the system once all inconsistencies have been worked out of the social relations---the divergence itself can be viewed as taking place beyond the window of time over which the system corresponds to anything real.  Alternately, one can imagine that as the community completes its separation into two groups, other social processes take over.  For example, individuals with differing ideological views or social preferences may self-segregate, breaking the all-to-all assumption of the model.  In other cases, mounting tensions may erupt into violence, reflecting a sort of bound on the relationship intensity achievable for pairs of nodes in the network.

Lastly we can ask, rather speculatively, whether the model provides any hints on how to guide divided communities toward reconciliation (in the cases where this is a sensible goal).  The work presented here implies that the mean friendliness of the social network should be an important target for modulation.  This suggests one potential strategy:  (i) direct the attention of the social network away from its divided status, (ii) encourage the formation of friendships across the divide, and then (iii) bring the network back to the task of managing the issue that originally divided it, with the hope that the increase in mean friendliness will push the network toward the all-friends configuration.  Remarkably Pettigrew, a social psychologist, has recently proposed a similar hypothesis with respect to overcoming prejudice, recommending the longitudinal process of (i) diverting attention away from ingroup-outgroup distinctions, (ii) allowing strong intergroup friendships to form, and then (iii) refocusing the community on social categorization until a single group category emerges \cite{pettigrew98,pettigrew06}.  Considering the differences in discipline and methodology, the similarity between  Pettigrew's sequence of steps and ours is striking, and the combined lesson is clear:  given the right combination of diversion and bonding exercises, it may be possible to get a fractured social network to resolve its differences and begin to heal.

\vspace{4 pt}

\textbf{Acknowledgments.}  Research supported in part by the John D. and Catherine T. MacArthur Foundation, a Google Research Grant, a Yahoo!~Research Alliance Grant, an Alfred P. Sloan Foundation Fellowship, a Microsoft Research New Faculty Fellowship, a grant from the Air Force Office of Scientific Research, and NSF grants CCF-0325453, BCS-0537606, CCF-0643934, IIS-0705774, and CISE-0835706.  We would also like to thank Nick Trefethen for pointers to the literature on matrix Riccati equations.

\end{document}


\title{Supporting Text}

\author{Seth A. Marvel\footnotemark[2]}

\author{Jon M. Kleinberg\footnotemark[3]}
\email{kleinber@cs.cornell.edu}

\author{Robert D. Kleinberg\footnotemark[3]}

\author{Steven H. Strogatz\footnotemark[2]}

\affiliation{\vspace{6 pt} \footnotemark[2]Center for Applied Mathematics, Cornell University, Ithaca, New York 14853 \vspace{1 pt} \\
\footnotemark[3]Department of Computer Science, Cornell University, Ithaca, New York 14853}
\maketitle

\section{Referenced Results}

In the introduction of our paper, we assert that a variant of the dynamics proposed by Ku{\l}akowski et al. generically balances an isolated triangle.  We explain what we mean here.

\vspace{10pt}

\textbf{Theorem 1.}  The system $\dot{x}_{12} = x_{13} x_{23}$, $\dot{x}_{13} = x_{12} x_{23}$, $\dot{x}_{23} = x_{12} x_{13}$ achieves balance when the initial values $x_{12}(0)$, $x_{13}(0)$ and $x_{23}(0)$ are all unequal.

\vspace{10pt}

\textbf{Proof.}  Multiplying each $\dot{x}_{ij}$ by $x_{ij}$ yields $x_{12} \dot{x}_{12} = x_{13} \dot{x}_{13} = x_{23} \dot{x}_{23}$.  Integrating these equalities gives the constraints $x_{12}^2 - x_{13}^2 = C_1$ and $x_{12}^2 - x_{23}^2 = C_2$ which partition the three-dimensional space of $(x_{12}, x_{13}, x_{23})$ into trajectories (with the direction of flow given by the original dynamical system).  Examination of this flow reveals that each initial condition $(x_{12}(0), x_{13}(0), x_{23}(0))$ with distinct coordinates flows into one of the four octants on which Heider balance holds, that is where $x_{12} x_{13} x_{23} > 0$.  Furthermore, these octants each act as separate trapping regions:  once a trajectory enters, it cannot leave.  Hence, the theorem follows. $\square$

\vspace{10pt}

The next theorem regards the main system of the paper with a rescaling of time:  $\frac{d}{dt}X = n^{-1} X^2$, where $X$ is a real symmetric $n \times n$ matrix.  Recall that $x_{ij}(t)$ denotes the $(i,j)$th element of the solution matrix $X(t)$ subject to the initial condition $X(0)$.  In the following, we will abbreviate $X(0)$ as $A$ and $x_{ij}(0)$ as $a_{ij}$.  Suppose that the $a_{ij}$, $i \leq j$, are drawn independently from distributions $F_{ij}$ with zero mass outside $[-K, K]$, and the off-diagonal distributions $F_{ij}$ have common expectation $\mu$ and variance $\sigma^2$.

\vspace{10pt}

\textbf{Theorem 2.}  $\lim_{n \rightarrow \infty} x_{ij}(t) = a_{ij} - \mu + \mu/(1 - \mu t)$ with probability 1 for $t \in [0, 1/K)$.

\vspace{10pt}

\textbf{Proof.}  Regard each step of the limit $n \rightarrow \infty$ as a selection and concatenation of elements $\{a_{in}\}_{1 \leq i \leq n-1}, \{a_{nj}\}_{1 \leq j \leq n-1}, a_{nn}$ to the elements $\{a_{ij}\}_{1 \leq i,j \leq n-1}$ selected in preceding steps.  Now consider the partial sum of the Taylor series expansion of $x_{ij}(t)$:
	\begin{equation} \label{17} 
	x_{ijnN}(t) = \sum_{k=0}^N \alpha_{kn} t^k \; \; \; \text{where \; $\alpha_{kn} 
	= \frac{1}{k!} \frac{d^k x_{ij}}{dt^k} \bigg|_{t=0}$}
	\end{equation}
The first step of the proof of Theorem 2 consists of proving that $\lim_{N \rightarrow \infty} \lim_{n \rightarrow \infty} x_{ijnN}(t)$ converges to $a_{ij} - \mu + \mu/(1 - \mu t)$ with probability 1 on $[0, 1/|\mu|)$ (see Lemma 1).  The second step of the proof consists of proving that $\lim_{N \rightarrow \infty} \lim_{n \rightarrow \infty} x_{ijnN}(t) = \lim_{n \rightarrow \infty} x_{ij}(t)$ with probability 1 on $[0, 1/K)$ (see Lemma 2).  Since we can write $\lim_{n \rightarrow \infty} x_{ij}(t)$ as $\lim_{n \rightarrow \infty} \lim_{N \rightarrow \infty} x_{ijnN}(t)$, this amounts to showing that the two limits can be exchanged on $[0, 1/K)$.  The above theorem then follows trivially by a union bound. $\square$

\vspace{10pt}

\textbf{Lemma 1.}  Under the assumptions of Theorem 2, $\lim_{N \rightarrow \infty} $ $\lim_{n \rightarrow \infty} x_{ijnN} = a_{ij} - \mu + \mu/(1 - \mu t)$ with probability 1 for $t \in [0, 1/|\mu|)$.

\vspace{10pt}

\textbf{Proof.}  For the sake of generality, we present a proof with more mild assumptions than those of the rest of the paper:  we only require that the moments of the $F_{ij}$ distributions be finite (and off-diagonal distributions to have mean $\mu$), not that the $a_{ij}$ values be bounded by $K$ with probability 1.  

Define $\alpha_{k\infty} = \lim_{n \rightarrow \infty} \alpha_{kn}$ (merely shorthand---we do not assume the limit exists).  By a union bound, we have
	\begin{equation} \label{18} 
	\Pr\bigl(\bigcap_{k=1}^\infty [\alpha_{k\infty} = \mu^{k+1}]\bigr) \geq 1 - \sum_{k = 1}^\infty [1 - \Pr(\alpha_{k\infty} = \mu^{k+1})]
	\end{equation}
So if we can show that $\Pr(\alpha_{k\infty} = \mu^{k+1}) = 1$ for all $k \geq 1$, then $\Pr(\bigcap_{k=1}^\infty [\alpha_{k\infty} = \mu^{k+1}]) = 1$.  In this case, $\lim_{N \rightarrow \infty} \lim_{n \rightarrow \infty}$ $x_{ijnN}$ has the convergent Taylor series $a_{ij} + \sum_{k=1}^\infty \mu^{k+1} t^k$ on $[0, 1/|\mu|)$ with probability 1, which proves the lemma.

So our task reduces to showing that $\Pr(\alpha_{k\infty} = \mu^{k+1}) = 1$ for each $k \geq 1$.  In order to do this, we need to compute the leading behavior of $\alpha_{kn}$ in $n$.  To calculate the $k$ time derivatives of $x_{ij}$ in the formula for $\alpha_{kn}$ (see Eq.~\textbf{\ref{17}}), we alternate between applying the chain rule of differential calculus and substituting in the right-hand side of $\dot{x}_{ij} = n^{-1} \sum_k x_{ik} x_{kj}$ (our system $\dot{X} = n^{-1} X^2$ written in element-wise fashion).  This gives
	\begin{equation} \label{19} 
	\alpha_{kn} = n^{-k} \sum_{m_1 = 1}^n \sum_{m_2 = 1}^n \dotsi 
	\sum_{m_k = 1}^n a_{i m_1} a_{m_1 m_2} \dotsm \, a_{m_k j}
	\end{equation}
where the factor $n^{-k}$ comes from the $k$ factors of $n^{-1}$ introduced by the $k$ derivatives, and the factor $1/k!$ in the formula for $\alpha_{kn}$ cancels with a factor $k!$ that arises from repeated applications of the chain rule.  In Eq.~\textbf{\ref{19}}, the dominant term is a sum of the edge value products of all simple length-$(k+1)$ paths between $i$ and $j$.  This sum contains $(n-2)!/(n-2-k)!$ terms.  All other paths include fewer immediate nodes and thus have at least a factor of $n$ fewer terms in their sums.

Our goal then for the remainder of the proof is to show that the first term of Eq.~\textbf{\ref{19}} is the only term that remains after taking $n$ to infinity, and that it converges to $\mu^{k+1}$ with probability 1.  To simplify notation, let $\ell$ denote the product of the edge values $a_{ij}$ along a particular path of length $k+1$ (not necessarily simple) from $i$ to $j$, and let $L$ denote the set of all such products on paths with the same configuration, or pattern of connectivity.  Denote the set of all $L$ by $\{L\}$, and let $S$ denote the one $L$ in $\{L\}$ consisting of simple paths of length $k+1$ from $i$ to $j$.  

Now observe that
$\cap_L{[\lim_{n \rightarrow \infty} n^{-k} \sum_{\ell \in L} \ell = 0]} \cap [n^{-k}$ $\sum_{\ell \in S} \ell = \mu^{k+1}] \subset [\alpha_{k\infty} = \mu^{k+1}]$.  So by another union bound,
	\begin{multline} \label{20}
	\Pr(\alpha_{k\infty} = \mu^{k+1}) 
	\geq 1 - \Pr\biggl(\lim_{n \rightarrow \infty} n^{-k} \sum_{\ell \in S} \ell \neq \mu^{k+1}\biggr) 
	\\ - \sum_{\{L\} \backslash S} \Pr\biggl(\lim_{n \rightarrow \infty} n^{-k} \sum_{\ell \in L} \ell \neq 0\biggr)
	\end{multline}
Hence, we are done if we can show that (i) $\Pr(\lim_{n \rightarrow \infty} n^{-k}$ $\sum_{\ell \in S} \ell = \mu^{k+1}) = 1$ for $S$ and (ii) $\Pr(\lim_{n \rightarrow \infty} n^{-k} \sum_{\ell \in L} \ell = 0) = 1$ for all other $L$.  Although $\sum_{\ell \in L} \ell$ is in general a sum of correlated random variables, it is possible to adapt a standard proof of the strong law of large numbers for uncorrelated random variables to prove both items.  We do this next.

Let's prove (ii) first.  For brevity, let $S_n = \sum_{\ell \in L} \ell$ and choose $v$ to denote the number of nodes in the path configuration of $L$.  For any positive $\epsilon$ and $r = 1, 2, \dotsc$, Markov's inequality gives $\Pr(|S_n| \geq (n\epsilon)^k) \leq E(|S_n|^r)/(n \epsilon)^{kr}$.  So if we can find an $r$ such that $E(|S_n|^r)/(n \epsilon)^{kr} \leq C/n^2$ for some constant $C$ (dependent on $\epsilon$), then $\sum_n \Pr(|S_n| \geq (n \epsilon)^k)$ converges, and by the first Borel-Cantelli lemma, $\Pr(|n^{-k} \sum_{\ell \in L} \ell| \geq \epsilon \text{ i.o.}) = 0$ for all $\epsilon > 0$ (where i.o.~stands for infinitely often).  Careful reflection reveals that $\cup_\epsilon [|n^{-k} \sum_{\ell \in L} \ell| \geq \epsilon \text{ i.o.}]$ (for, say, all rational $\epsilon$) is the complementary event of $[\lim_{n \rightarrow \infty} n^{-k} \sum_{\ell \in L} \ell = 0]$, and so we have arrived at the desired result (ii).

Hence, in order to actually show (ii), we need to find an $r$ such that $E(|S_n|^r)/(n \epsilon)^{kr} \leq C/n^2$.  Consider $r = 2$:  $E(S_n^2) = \sum E(\ell_x \ell_y)$, where each index of the sum ranges independently over $L$.  There are $(n-2)!/(n-2-v)!$ paths $\ell$ in $L$, so there are fewer than $n^{2v}$ terms in $\sum E(\ell_x \ell_y)$, and $E(S_n^2) \leq Dn^{2v}$ for some constant $D$.  Since $v < k$ for all $L$ other than $S$, we have $E(|S_n|^2)/(n \epsilon)^{2k} \leq  Dn^{2v}/(n \epsilon)^{2k} \leq C/n^2$ where $C = D\epsilon^{-2k}$, and the proof of (ii) is complete.

Finally, to prove (i), start by replacing each factor $a_{xy}$ in $\ell$ with $b_{xy} + \mu$, where $b_{xy} = a_{xy} - \mu$.  Now expand the result and cancel $\mu^{k+1}$ from both sides of $n^{-k} S_n = \mu^{k+1}$ to obtain $n^{-k} S_n' = 0$, where $S_n'$ is a sum over $S$ of a polynomial $Q$ with $2^{k+1}-1$ terms, each of the form $\mu\mu \dotsm \mu b_{uv} b_{wx} \dotsm b_{yz}$ where at least one of the factors is a $b_{xy}$ and the total number of factors in the term is $k+1$.  Note that each place of $Q$ corresponds to a particular set of $b_{xy}$'s from the original simple path, e.g. the 14th place of $Q$ might have $b_{xy}$'s corresponding to the 1st, 4th, 5th, and 7th edges of the path, and $\mu$'s for the other edges.  Now let $m_q$ denote the number of vertices (excluding $i$ and $j$) among the subscripts of the $b_{xy}$'s in a given term.  The remaining $k-m_q$ nodes of the path not found in the term (supplanted by the $\mu$'s) can take any of $(n - 2 - m_q)!/(n - 2 - k)!$ permutations.  Hence, there are no more than $n^{k-m_q}$ identical copies of any one term in $S_n'$ from the same place in $Q$.

Now consider one of the $(2^{k+1} - 1)^4$ ways that terms in the $2^{k+1} - 1$ places of $Q$ can be multiplied together in $S_n'^4$.  Note that this can produce no more than $n^{4k-\sum_{q=1}^4 m_q}$ identical copies of the same term.  Second, since the $b_{xy}$'s each have expectation zero, every $b_{xy}$ in the final term must appear to at least a power of two or the whole term has expectation zero.  This implies that for each nonvanishing term, there must be some pattern of matching between the $b_{xy}$'s.  The number of possible matchings is clearly a function of $k$ and not $n$ (it certainly is not more than the number of partitions of $4(k+1)$ edges), so consider one of these possible matchings.  Now observe that if, as we stated above, we consider only one of the $(2^{k+1} - 1)^4$ ways that terms in the $2^{k+1} - 1$ places of $Q$ can be multiplied together in $S_n'^4$, then no more than $n^{\sum_{q=1}^4 m_q/2}$ distinct nonvanishing terms can be constructed per matching for any such way of combining terms.  This holds because each $b_{xy}$ needs at least one match, and so the number of free nodes cannot exceed half the total number of $b_{xy}$'s in the final term.  Thus we have shown the highest order of $n$ possible for $E(S_n'^4)$ is given by the maximum value of $n^{\sum_{q=1}^4 m_q/2} n^{4k-\sum_{q=1}^4 m_q}$.  Since $m_q \geq 1$ for each $q = 1, \dotsc, 4$, this can at most be $n^{4k-2}$, which by the above reasoning completes the proof of (i) and hence the full theorem. $\square$

\vspace{10pt}

\textbf{Lemma 2.}  Under the assumptions of Theorem 2, $\lim_{n \rightarrow \infty}$ $\lim_{N \rightarrow \infty} x_{ijnN} = \lim_{N \rightarrow \infty} \lim_{n \rightarrow \infty} x_{ijnN}$ with probability 1 for $t \in [0, 1/K)$.

\vspace{10pt}

\textbf{Proof.}  We need three ingredients for this proof.  We will first describe the three ingredients and then show how they together prove Lemma 2.  Throughout the following, all statements hold with probability 1 unless stated otherwise.  

As we found in the course of the proof of Lemma 1, the limits $\lim_{n \rightarrow \infty} \alpha_{kn}$ exist for all $k$ and are $\mu^{k+1}$ on $[0, 1/|\mu|)$, so $\lim_{n \rightarrow \infty} \sum_{k=0}^N \alpha_{kn}t^k$ exists under the same conditions, and we call it $x_{ij\infty N}(t)$.  This gives us the first ingredient:  (i) $\lim_{n \rightarrow \infty} x_{ijnN}(t) = x_{ij\infty N}(t)$ for $t \in [0, 1/|\mu|)$ and any $N$.  Additionally, from Lemma 1 we know that $\lim_{N \rightarrow \infty} x_{ij\infty N}(t)$ exists and is $a_{ij} - \mu + \mu/(1 - \mu t)$ on $[0, 1/|\mu|)$.  We call this limit $x_{ij\infty\infty}(t)$, and write the second ingredient as (ii) $\lim_{N \rightarrow \infty} x_{ij\infty N}(t) = x_{ij\infty\infty}(t)$ for $t \in [0, 1/|\mu|)$.

Finally, as we saw in the proof of Lemma 1, $\alpha_{kn} = n^{-k} \sum a_{i m_1} a_{m_1 m_2} \dotsm \, a_{m_k j}$ (by definition, not just with probability 1), where the $k$ indices $m_x$ each range independently from $1$ to $n$.  Since each $|a_{ij}| < K$, we must have that $|\alpha_{kn}| \leq K^{k+1}$, which implies $|x_{ijn\infty}(t) - x_{ijnN}(t)| \leq K(Kt)^{N+1}/(1-Kt)$.  So if $|Kt| < 1$, then for any $\epsilon > 0$, there is a sufficiently large $N_1$ independent of $n$ such that $|x_{ijn\infty}(t) - x_{ijnN}(t)| \leq \epsilon$ for all $N \geq N_1$.  This constitutes our third ingredient, that $x_{ijnN}(t)$ converges uniformly to $x_{ijn\infty}(t)$:  (iii) for every $\epsilon > 0$, there exists an $N_1$ such that for all $N \geq N_1$ and all $n$, $|x_{ijn\infty}(t) - x_{ijnN}(t)| < \epsilon$.

To complete the proof of Lemma 2, we need to show that $\lim_{n \rightarrow \infty} x_{ijn\infty}(t)$ exists and is just $x_{ij\infty\infty}(t)$ on $[0, 1/K)$.  Start by picking an $\epsilon > 0$.  Then by (iii), there exists an $N_1$ such that if $N > N_1$ then $|x_{ijn\infty}(t) - x_{ijnN}(t)| < \epsilon$ for all $n$.  Similarly, (ii) implies that there exists an $N_2$ such that if $N \geq N_2$, then $|x_{ij\infty\infty}(t) - x_{ij\infty N}(t)| < \epsilon$.  Finally, let $N_3 = \max\{N_1,N_2\}$.  Then by (i), we may choose an $n_1$ such that if $n \geq n_1$, then
$|x_{ij\infty N_3}(t) - x_{ijnN_3}(t)| < \epsilon$.  Now define the following events:
	\begin{equation} \label{21}
	\begin{split}
	E_1 &= [|x_{ij\infty\infty}(t) - x_{ij\infty N_3}(t)| < \epsilon] \\ 
	E_2 &= [|x_{ij\infty N_3}(t) - x_{ijnN_3}(t)| < \epsilon] \\
	E_3 &= [|x_{ijnN_3}(t) - x_{ijn\infty}(t)| < \epsilon] \\
	E_4 &= [|x_{ij\infty\infty}(t) - x_{ijn\infty}(t)| < 3\epsilon]
	\end{split}
	\end{equation}
Observe that, in similar form to Eq.~\textbf{\ref{20}}, $(E_1 \cap E_2 \cap E_3) \subset E_4$, so $\Pr(E_4) \geq \Pr(E_1 \cap E_2 \cap E_3) = 1 - \Pr(E_1' \cup E_2' \cup E_3') \geq 1 - \Pr(E_1') - \Pr(E_2') - \Pr(E_3') = 1$ for all $n \geq n_1$.  Thus, $|x_{ij\infty\infty}(t) - x_{ijn\infty}(t)| < 3\epsilon$ for all $n \geq n_1$ and $t \in [0, 1/K)$. $\square$

\section{Unreferenced Results}

In the main report, we establish that $\lambda_1 > 0$ in probability by way of Wigner's semicircle law.  However, we can also show that $\lambda_1 > 0$ with high probability under a different set of assumptions about how $A$ is selected.  Loosely speaking, the selection of the diagonal entries is more constrained in this alternative approach while the selection of the off-diagonal entries is less so.

Suppose that all $a_{ij}$, $|i - j| = 1$, are chosen from the same distribution $F$ and all $a_{ii}$ are chosen from the same distribution $G$.  All selections are independent for $i \leq j$.  In addition, assume that the density of $a_{11} - a_{12}$ is not entirely confined to negative values.  Then we have the following theorem.

\vspace{10pt}

\textbf{Theorem 3.}  $\Pr(\lambda_1 \leq 0)$ is exponentially small in $n$.

\vspace{10pt}

\textbf{Proof.}  If $\lambda_1 \leq 0$, then the corresponding matrix $A$ is negative semi-definite.  Therefore $v^T A v \leq 0$ for every vector $v$ ($T$ denotes transposition).  Let $v_k$ denote the vector with $+1$ and $-1$ in its $(2k-1)$th and $2k$th components, respectively, and $0$ in all other components.  Then the event that $v_k^T A v_k \leq 0$ is equivalent to the event that $a_{(2k-1)(2k-1)} - a_{(2k-1)2k} - a_{2k(2k-1)} + a_{2k2k} \leq 0$.  Note that the left-hand side of this final inequality has at least a constant probability of being positive.  Now as $k$ ranges from $1$ to $n/2$, we encounter $n/2$ independent events, each having at least a constant probability of failure.  Hence the probability that $A$ is negative semi-definite is exponentially small in $n$. $\square$

\vspace{10pt}

Very roughly, the final result of this supporting text says that in the case of negative $\mu$, the distribution of the \textit{sum} of the components of $\omega_1 = (v_1, \cdots, v_n)$ retains no more than constant width in the large-$n$ limit.  So, for example, the mean of the components of $\omega_1$ must shrink to zero at least as fast as $1/n$.  (Note that this convergence is faster than the $1/\sqrt{n}$ convergence of means for independent and identically distributed random variables with finite mean and variance.)  This result is consistent with the picture that when $\mu$ is negative, the system is destined for two-sided conflict in the large-$n$ limit.

To make the proof less cumbersome, suppose that the diagonal $F_{ii}$ have common expectation $\mu$ and variance $\sigma^2$ just like the off-diagonal $F_{ij}$.

\vspace{10pt}

\textbf{Theorem 4.}  If $\mu < 0$, then $\lim_{n \rightarrow \infty} \Pr(|n^{-\delta}$ $\sum_{i=1}^n v_i| < \epsilon) = 1$ for all $\delta, \epsilon > 0$.

\vspace{10pt}

\textbf{Proof.}  Consider the eigenvalue equation $\sum_{j=1}^n a_{ij} v_j = \lambda_1 v_i$.  Sum both sides over $i$, and then rearrange and switch index labels to obtain
	\begin{equation} \label{12}
	\sum_{i=1}^n v_i \sum_{j=1}^n a_{ij} = \lambda_1 \sum_{i=1}^n v_i.
	\end{equation}
We anticipate from standard results of random matrix theory that the left side of Eq.~\textbf{\ref{12}} is asymptotic to $\mu n \sum_{i=1}^n v_i$ and the right side to $2\sigma\sqrt{n} \sum_{i=1}^n v_i$.  So let's define the two events
	\begin{equation} \label{13}
	\begin{split}
	E_1 &= \biggl[ \Bigl| n^{-1-\delta} \sum_{i=1}^n v_i \sum_{j=1}^n a_{ij} 
	- \frac{\mu}{n^\delta} \sum_{i=1}^n v_i \Bigr| \leq \epsilon \biggr] \\ 
	E_2 &= \biggl[ \Bigl| \frac{\lambda_1}{n} \sum_{i=1}^n v_i 
	- \frac{2 \sigma}{\sqrt{n}} \sum_{i=1}^n v_i \Bigr| < \epsilon \biggr]
	\end{split}
	\end{equation}
where $[\cdot]$ denotes an event.  Now, $\Pr(E_1 \cap E_2) = 1 - \Pr(E_1' \cup E_2') \geq 1 - \Pr(E_1') - \Pr(E_2')$, where we have written $E_x'$ for the complementary event of $E_x$.  So if we can show that $\lim_{n \rightarrow \infty} \Pr(E_1) = 1$ and $\lim_{n \rightarrow \infty} \Pr(E_2) = 1$ for all $\delta, \epsilon > 0$, it follows that $\lim_{n \rightarrow \infty} \Pr(|\mu n^{-\delta} \sum_{i=1}^n v_i - 2 \sigma n^{-1/2-\delta} \sum_{i=1}^n v_i| < \epsilon + \epsilon n^{-\delta}) = 1$, which implies Theorem 4.

Hence our task reduces to showing that $\lim_{n \rightarrow \infty} \Pr(E_1) = \lim_{n \rightarrow \infty} \Pr(E_2) = 1$.  First consider $\Pr(E_1)$.  By Jensen's inequality and normalization of $\omega_1$, $(n^{-1} \sum_{i=1}^n |v_i|)^2 \leq n^{-1} \sum_{i=1}^n v_i^2 = n^{-1}$, so 
	\begin{equation} \label{14}
	\biggl| \sum_{i=1}^n v_i \biggr| \leq \sum_{i=1}^n |v_i| \leq \sqrt{n}.
	\end{equation}
Now define the additional event:
	\begin{equation} \label{15}
	E_3 = \biggl[\max_{1 \leq i \leq n} \Bigl| n^{-1/2-\delta} 
	\sum_{j=1}^n (a_{ij} - \mu) \Bigr| \leq \epsilon \biggr].
	\end{equation}
By Eq.~\textbf{\ref{14}}, we have $E_3 \subset [\max_{1 \leq i \leq n} |n^{-1/2-\delta} \sum_{j=1}^n (a_{ij} - \mu)| n^{-1/2} \sum_{i=1}^n |v_i| \leq \epsilon]
\subset [n^{-1-\delta} \sum_{j=1}^n |v_i| |\sum_{j=1}^n (a_{ij} - \mu)| \leq \epsilon]
\subset E_1$.  The Bernstein inequality and a union bound together imply that $\lim_{n \rightarrow \infty} \Pr(E_3) = 1$, since they give that
	\begin{equation} \label{16}
	\Pr(E_3') \leq 2n\exp\left(-\frac{n^{2\delta} \epsilon^2/2}
	{\sigma^2+(K+\mu)n^{-1/2+\delta}\epsilon/3}\right).
	\end{equation}
So, $\lim_{n \rightarrow \infty} \Pr(E_1) = 1$ as desired.

Regarding $\Pr(E_2)$, we know that $\lambda_1 \in 2\sigma\sqrt{n} + o(\sqrt{n})$ in probability~\cite{furedi81}, which implies that $\lim_{n \rightarrow \infty} \Pr(|\lambda_1/\sqrt{n} - 2 \sigma| < \epsilon) = 1$ and therefore $\lim_{n \rightarrow \infty} \Pr(E_2) = 1$ for all $\epsilon > 0$ by Eq.~\textbf{\ref{14}}.  This completes the proof. $\square$

\section{Additional References}

Examples of two-sided social conflicts appear in a wide range of settings, including political party coalitions~\cite{duverger54,riker62,riker82,benoit07,sagar09}, inter-state wars~\cite{moore79,altfeld79}, corporate standard-setting~\cite{axelrod95}, intertribal feuding~\cite{chagnon88} and even experiments in which group membership is based on arbitrary criteria~\cite{sherif66,tajfel79,brewer79,fiske02,wildschut03}.